\documentclass[twocolumn]{aastex61}
\received{..........}
\revised{..........}
\accepted{\today}
\submitjournal{ApJ}

\shorttitle{Abundances in FSR 1758}
\shortauthors{Villanova et al.}

\begin{document}

\title{Detailed chemical composition and orbit of the new globular cluster FSR1758: Implications for the accretion of the Sequoia dwarf galaxy onto the Milky Way}\thanks{Based on observations carried out at .............
under programs .............}

\correspondingauthor{Sandro Villanova}
\email{svillanova@astro-udec.cl}

\author{Sandro Villanova}
\affiliation{Departamento de Astronomia, Casilla 160-C, Universidad de Concepcion, Cooncepci{\'o}n, Chile}

\author{Lorenzo Monaco}
\affiliation{Departamento de Ciencias Fisicas, Facultad de Ciencias Exactas, Universidad Andres Bello, Av. Fernandez Concha 700, Las Condes, Santiago, Chile}
\nocollaboration

\author{Julia O'Connell}
\affiliation{Departamento de Astronomia, Casilla 160-C, Universidad de Concepcion, Cooncepci{\'o}n, Chile}

\author{Doug Geisler}
\affiliation{Departamento de Astronomia, Casilla 160-C, Universidad de Concepcion, Cooncepci{\'o}n, Chile}
\nocollaboration

\author{Dante Minniti}
\affiliation{Departamento de Ciencias Fisicas, Facultad de Ciencias Exactas, Universidad Andres Bello, Av. Fernandez Concha 700, Las Condes, Santiago, Chile}
\affiliation{Millennium Institute of Astrophysics, Av. Vicuna Mackenna 4860, 782-0436, Santiago, Chile}
\affiliation{Vatican Observatory, V00120 Vatican City State, Italy}

\author{Paulina Assmann}
\affiliation{Departamento de Astronomia, Casilla 160-C, Universidad de Concepcion, Cooncepci{\'o}n, Chile}

\author{Rodolfo Barb{\'a}}
\affiliation{Departamento de Fisica y Astronomia, Universidad de La Serena, Avenida Juan Cisternaa 1200, La Serena, Chile.}

\nocollaboration



\begin{abstract}
We present detailed chemical abundances, radial velocities and orbital parameters for FSR 1758, a recently discovered star cluster in the direction of the Galactic Bulge. High resolution (R$\sim$42,000) spectra were obtained using the Magellan/Clay telescope instrumented with MIKE echelle spectrogragh, wavelength range $\sim$4900-8700 \AA. Cluster membership was determined using Gaia DR2 proper motions and confirmed with our radial velocity measurements. We find metallicity consistent with previous photometric estimates for this cluster, [Fe/H] = -1.58$\pm$0.03 dex, with a small, 0.08 dex, spread. While other studies have suggested this massive object may be the result of a previous accretion event, our results are consistent with Milky Way Halo globular clusters with characteristic Na-O anti-correlations found for the metal-poor cluster members. The mean radial velocity of the cluster, +226.8$\pm$1.6 km s$^{\rm -1}$ with a small velocity dispersion, 4.9$\pm$1.2 km s$^{\rm -1}$, is typical for globular clusters. We also confirm a retrograde Galactic orbit that appears to be highly eccentric.
\end{abstract}




\section{Introduction} 

The census of Galactic globular clusters (GCs) appears to be incomplete in the direction of the central regions of the Milky Way due to high interstellar absorption and source crowding ( e.g. \citealt{Iv05,Iv17}), as illustrated by several recent discoveries 
(e.g. \citealt{Mi17a,Mi17b,Ry18,Bor18,Ca18,Ca19,Pa19}).
While most of the new GC candidates have low luminosity, one of the most unexpected discoveries is FSR 1758, that is a large metal-poor GC recently found in the direction of the Galactic bulge \citep{Ba19}, centered at coordinates $RA=17:31:12$, and $DEC= -39:48:30$ (J2000), and Galactic coordinates $l = 349.217$ deg, $b = -3.292$ deg.
They use data from Gaia, the DECaPS optical survey \citep{Sc18} and the VVVX survey
\citep{Mi10,Mi18} to estimate the GC parameters.
In particular, Barba et al. (2019) determined a metallicity $[Fe/H]=-1.5 \pm 0.3$ dex based on the optical and near-IR CMDs from Gaia and VVV, respectively.
They also measure a distance $D=11.5 \pm 1.0$ kpc, reddening $E(J-Ks)=0.20 \pm 0.03$ mag, radii $Rc =10$ pc, and $Rt =150$ pc, total magnitude $M_i \sim -8.6 \pm 1.0$ mag, and proper motions $\mu_\alpha = -2.95$ mas/yr, and $\mu_\delta = 2.55$ mas/yr. 

The determination of the fundamental parameters of this new GC is important because it has been recently proposed that FSR 1758 belongs to the past accretion of a dwarf galaxy (named Sequoia) by the Milky Way, along with other GCs \citep{My19}.
These authors argue that the Sequoia event along with the so called Sausage event are the two most significant accretion events that occurred in the Milky Way.

\citet{Ba19} argues that FSR 1758 may be part of an extended dwarf galaxy disrupted by the Milky Way, along with possibly Ton2 and NGC 6380, that are located within 1 deg of FSR 1758. However, the RVs of these clusters ($RV_{Ton2}=-182$ km/s and $RV_{NGC6380}=-4$ km/s) are very different from that of FSR 1758 ($RV_{FSR1758}=+227 km/s$), making the association very unlikely, based on the Gaia RV measurement of \citet{Si19},  who concludes that this cluster has a retrograde orbit.

On the other hand, based on the strongly retrograde orbit, \citet{My19} argues that it belongs to the Sequoia dwarf galaxy, accreted by the Milky Way along with the GCs WCen, NGC3201, NGC6101, NGC6535, NGC6388, and NGC6401. They argue that these clusters share common properties in the age-metallicity space for example.

As a caveat, according to \citet{My19}, the Gaia-Enceladus event \citep{He18} includes both the Sequoia and Sausage events. Much remains to be understood in terms of the past Milky Way accretion events, and globular clusters with well determined physical parameters, like is now FSR 1758, are key elements of evidence.

We have obtained high dispersion spectra for 9 stars that are confirmed members of the cluster. The deep photometry and accurate proper motions measured by Gaia and VVV are complemented by our high dispersion spectroscopy to fully characterize this GC, defining its parameters.
The spectroscopy presented in this paper allows us to measure accurate radial velocities and detailed chemical abundances for the target stars.
In particular, the Gaia proper motions in combination with our accurate radial velocities are employed to improve the cluster's orbit. 

In this paper we aim to answer the following questions:
What is the detailed chemical composition of the new GC FSR 1758?
Does it have a measurable dispersion in chemical elements, suggesting the presence of multiple populations? How massive is this cluster? Is this a real GC or the nucleus of an accreted dwarf? Does this cluster belong to Milky Way GCS or to an external dwarf that has been accreted? What is the cluster orbit? Is this a halo, bulge, or disk GC?

\begin{figure}
\plotone{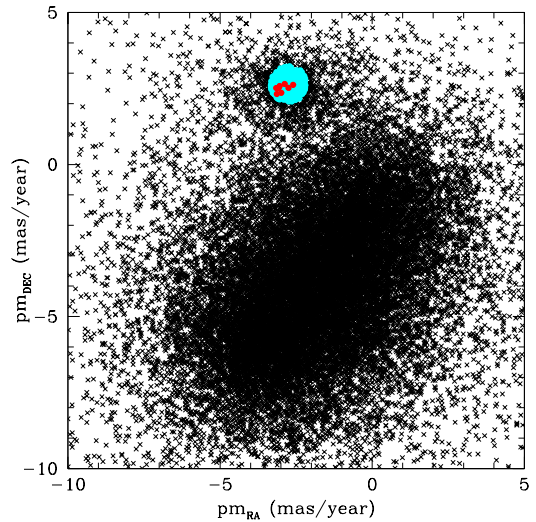}
\caption{Proper motions in the field of FSR 1758. Cluster members are the
  indicated with cyan circles, while our targets with red points.}
\label{fig01}
\end{figure}

\begin{figure}
\plotone{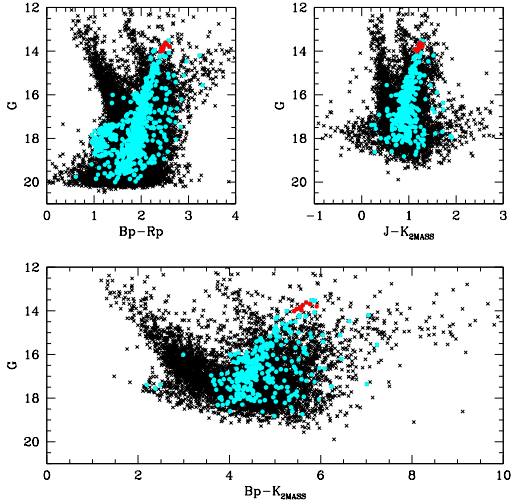}
\caption{CMDs of the member stars (cyan circles). Field stars are indicated
  with black crosses, while our targets are the red circles.}
\label{fig02}
\end{figure}

\section{Observations and data reduction}

Our data set consists of high-resolution spectra collected in 2018 October-November. The spectra come from single 300s exposures, obtained with the MIKE slit echelle spectrograph at the Magellan/Clay telescope. Weather conditions were excellent with a typical seeing of 0.7 arcsec at the zenith and a clear sky. We selected 9 isolated stars around the RGB tip. The MIKE spectra have a spectral coverage from 4900 to 8700 
\AA\ with a resolution of $\sim$42,000 and signal-to-noise ratio (S/N)$\sim$40 at 6300 \AA. 
Data were reduced using MIKE dedicated pipeline \citep{Ke00,Ke03} \footnote{ https://code.obs.carnegiescience.edu/mike}, including bias subtraction, flat-field correction, and wavelength calibration, scattered-light and sky subtraction.  Single orders were rectified and combined using IRAF \footnote{IRAF is distributed by the National Optical Astronomy Observatory, which is operated by the Association of Universities for Research in Astronomy, Inc., under cooperative agreement with the National Science Foundation.} .

In order to select our targets we took advantage of the proper motions published in the
Gaia DR2 \citep{Ga18} since the cluster suffers of a severe foreground contamination. Proper motions of the cluster field stars are reported in Fig.~\ref{fig01}, where cluster members are indicated with cyan symbols. As members we selected those stars within a radius of 0.7 mas/year around the center of the cluster in the proper motion space. The value of the radius was determined empirically assuming 1 mas/year as initial guess, and reducing it as much as possible in order to remove field contamination as shown from the CMD. We found that reducing the radius below 0.7 mas/year did not improve the CMD significantly.
Among these members we selected our targets, indicated by red circles.
The targets are RGB stars close to the tip. The mean proper motions of the members are:\\

pm$_{RA}$=-2.7911$\pm$0.0097\ mas/year\\

pm$_{DEC}$=+2.6042$\pm$0.0090\ mas/year\\

Gaia photometry was cross-correlated with 2MASS and we present multi-band CMDs of the cluster in Fig.~\ref{fig02}. The cluster presents well populated RGB and HB and shows to be affected by differential reddening. The HB is not visible in the infrared bands since HB stars emit most of their flux in the blue and ultraviolet.
Radial velocities were measured with the {\it fxcor} package in IRAF, using a synthetic spectrum as a template.  The mean radial velocity we obtained is 226.8$\pm$1.6 km/s.  There are no clear outliers in the radial velocity
distribution. If we consider also that all our targets have the same
proper motions (see Table~\ref{t1}) we conclude that they are all
cluster members. We found a radial velocity dispersion of 4.9$\pm$1.2 km/s
Table \ref{t1} lists the basic parameters of the 9 observed targets:
ID, J2000.0 coordinates, proper motions and G, B$_{P}$, R$_{P}$ magnitudes
from Gaia Data release 2, and J, H, K$_s$ magnitudes from 2MASS.

\begin{table*}
\tabcolsep 0.05truecm
\caption{ID, coordinates, proper motions and magnitudes of the observed
  stars. See text for details.}
\begin{tabular}{lcccccccccc}
\hline
ID & RA(2000.0) &DEC(2000.0) & pm$_{RA}$ & pm$_{DEC}$ & G & B$_{P}$ & R$_{P}$ &J &H &K\\
\hline
 & [hours] & [deg] & [mas/yr] & [mas/yr] & mag & mag & mag & mag & mag & mag \\
\hline\hline
01  & 17:31:08.7 & -39:49:42.2 &  -2.99056 & 2.35811 & 13.688 & 15.069 & 12.529 & 10.587 &  9.570 & 9.299\\ 
02  & 17:31:49.3 & -39:56:16.5 &  -2.87837 & 2.64815 & 13.797 & 15.227 & 12.629 & 10.604 &  9.577 & 9.315\\ 
03  & 17:31:17.3 & -39:50:30.5 &  -3.16035 & 2.51542 & 13.831 & 15.159 & 12.688 & 10.814 &  9.828 & 9.586\\ 
04  & 17:31:20.8 & -39:45:48.1 &  -3.04149 & 2.58301 & 13.866 & 15.161 & 12.736 & 10.821 &  9.873 & 9.617\\ 
05  & 17:30:56.0 & -39:50:38.5 &  -2.59973 & 2.61964 & 13.914 & 15.176 & 12.719 & 10.829 &  9.968 & 9.702\\ 
06  & 17:31:10.7 & -39:50:30.6 &  -2.75174 & 2.51828 & 14.016 & 15.296 & 12.888 & 11.057 & 10.111 & 9.878\\ 
045 & 17:31:10.8 & -39:47:02.8 &  -3.13759 & 2.31699 & 13.616 & 14.971 & 12.456 & 10.525 &  9.551 & 9.289\\ 
087 & 17:31:21.6 & -39:50:12.2 &  -3.07022 & 2.52644 & 13.971 & 15.278 & 12.822 & 10.896 &  9.948 & 9.700\\ 
097 & 17:31:28.7 & -39:54:06.2 &  -3.09229 & 2.51471 & 13.735 & 15.059 & 12.587 & 10.665 &  9.735 & 9.451\\
\hline\hline
\end{tabular}
\label{t1}
\end{table*}

\begin{table*}
\tabcolsep 0.05truecm
\caption{Parameters and abundances of the observed stars. The last rows give the
  mean abundances of the cluster and the relative error from the mean.}
\begin{tabular}{lcccccccccccccccc}
\hline
ID & $\rm T_{eff}$ &log(g) & $v_t$ & RV$_{HELIO} $&[Fe/H]&[O/Fe]&[Na/Fe]&[Mg/Fe]&[Al/Fe]&[Ca/Fe]&[TiI/Fe]&[TiII/Fe]&[Cr/Fe]&[Ni/Fe]&[Ba/Fe]&[Eu/Fe]\\
\hline
& $\rm ^oK$ & &[km/s] &[km/s] & dex & dex & dex & dex & dex & dex & dex & dex & dex & dex & dex & dex\\ 
\hline\hline
01  & 3880 &  0.00 & 1.83 & 229.6 &-1.52 & 0.15 & 0.25 & 0.62 & 0.45 & 0.40 & 0.39 & 0.44 &-0.18 &-0.16 & 0.12 & 0.25\\  
02  & 3880 &  0.00 & 1.91 & 223.8 &-1.55 &-0.06 & 0.61 & 0.74 & 0.77 & 0.42 & 0.34 &  -   &-0.33 &-0.05 & 0.07 & 0.28\\
03  & 4020 &  0.25 & 1.95 & 229.7 &-1.62 & 0.29 & 0.05 & 0.69 & 0.41 & 0.37 & 0.33 & 0.42 &-0.27 &-0.04 &-0.17 & 0.36\\
04  & 4010 &  0.28 & 1.88 & 226.2 &-1.54 & 0.13 & 0.49 & 0.66 & 0.40 & 0.37 & 0.37 & 0.28 &-0.22 &-0.08 &-0.13 & 0.19\\
05  & 4170 &  0.09 & 2.30 & 217.5 &-1.77 & 0.27 & 0.01 & 0.74 &  -   & 0.17 & 0.08 & 0.38 &-0.29 &-0.14 & 0.08 & 0.19\\
06  & 4080 &  0.43 & 1.72 & 234.7 &-1.57 & 0.32 &-0.02 & 0.60 & 0.49 & 0.34 & 0.34 & 0.31 &-0.18 &-0.13 & 0.10 & 0.46\\
045 & 3870 &  0.15 & 1.93 & 228.5 &-1.64 & 0.31 & 0.22 & 0.82 & 0.17 & 0.19 & 0.24 &  -   &-0.31 &-0.05 &-0.05 & 0.39\\
087 & 4035 &  0.58 & 1.92 & 222.9 &-1.52 & 0.25 & 0.24 & 0.55 & 0.46 & 0.34 & 0.39 & 0.25 &-0.20 &-0.03 &-0.15 & 0.48\\
097 & 3950 &  0.10 & 1.85 & 228.1 &-1.52 & 0.16 & 0.33 & 0.79 & 0.21 & 0.31 & 0.38 & 0.29 &-0.23 &-0.13 &-0.05 & 0.24\\
\hline
Mean  &    &      &      &  226.8 &-1.58 & 0.20 & 0.24 & 0.69 & 0.42 & 0.32 & 0.32 & 0.34 &-0.25 &-0.09 &-0.02 &0.32 \\
Error &    &      &      &    1.6 & 0.03 & 0.04 & 0.07 & 0.03 & 0.07 & 0.03 & 0.03 & 0.03 & 0.02 & 0.02 & 0.04 & 0.04\\
\end{tabular}
\label{t2}
\end{table*}

\section{Abundance analysis}

Figure ~\ref{fig02} shows that the cluster is affected by differential reddening.
For this reason an estimation of stellar parameters bases on photometry would require
a differential reddening correction. Therefore as initial temperature, gravity and microturbulence 
for our stars we just assumed tipical values for RGB stars close to the tip, that is T$_{\rm eff}$=4000 K, 
log(g)=0.5 dex, and v$_{\rm t}$=1.90 km/s.
Atmospheric models were calculated using ATLAS9 code \citep{Ku70},
assuming our estimations of T$_{\rm eff}$, log(g), and v$_{\rm t}$, and the
[Fe/H] value from \citet{Ba19}.\\ 
Then T$_{\rm eff}$, log(g), and v$_{\rm t}$ were re-adjusted and new 
atmospheric models calculated in an interactive way in order to remove trends 
in excitation potential and reduced equivalent width (EQW) versus abundance for T$_{\rm eff}$ and v$_{\rm t}$, respectively, 
and to satisfy the ionization equilibrium for log(g). 60-70 FeI lines and
 6-7 FeII lines (depending on the S/N of the spectrum) were used for the
latter purpose. The [Fe/H] value of the model was changed at each 
iteration according to the output of the abundance analysis. 
The Local Thermodynamic Equilibrium (LTE) program MOOG \citep{Sn73} was used
for the abundance analysis.

CaI, TiI, TiII, CrI, FeI, FeII, and NiI abundances were estimated
using the EQW method. For this purpose we measured EQW using the automatic
program DAOSPEC \citep{St08} \footnote{DAOSPEC is freely distributed by http://www.cadc-ccda.hia-iha.nrc-cnrc.gc.ca/en/community/STETSON/daospec/}. 
OI, NaI, MgI, AlI, BaII, and EuII abundances were obtained using the spectro-synthesis method. For this
purpose 5 synthetic spectra were generated for each line with 0.25 dex
abundance step and compared with the observed spectrum. The 
line-list and the methodology we used are the same used in previous papers (e.g. \citealt{Vi13}),
so we refer to those articles for a detailed discussion about this point. 
Here we just underline that we took hyperfine splitting into account for Ba
as in our previous studies. This is particularly important because Ba
lines are very strong even in metal-poor stars and hyperfine splitting help to
remove the line-core saturation producing a change in the final abundance as
estimated by the spectro-synthesis method up to 0.1 dex. Also  Eu  has odd-isotopes
affected by hyperfine splitting, but their lines are weak and the
line-core saturation is not at work. So hyperfine splitting corrections are negligible.

The abundances we obtained are reported in Tab.~\ref{t2}together with the mean values for the cluster and the error on the mean. For Ti we reported the mean values of TiI and TiII abundances.
Na is an element affected by NLTE effets. For this reason we looked in
the INSPEC \footnote{version 1.0 (http://inspect.coolstars19.com/index.php?n=Main.HomePage)} 
database for suitable NLTE corrections. We found the they are very small
($\sim$-0.05 dex) with no significant variation (less then 0.02 dex) in our
temperature range. For this reason we decided not to apply them to our Na abundances.

\begin{table*}
\tabcolsep 0.05truecm
\caption{Estimated errors on abundances due to errors on atmospheric
parameters and to spectral noise for star \#097 (column 2 to 6). Column 7
gives the total error calculated as the root squared of the sum of the squared
of columns 2 to 6. This total error must be compared with the observed
dispersion (RMS) of the data with its error (column 8). The last column gives the significance of the difference between
the total error and the observed dispersion, in units of $\sigma$. See text for more details.}     
\begin{tabular}{lccccccccc}        
\hline\hline  
El. & $\Delta$T$_{\rm eff}$=40 K  & $\Delta$log(g)=0.15 & $\Delta$v$_{\rm t}$=0.05 km/s
& $\Delta$[Fe/H]=0.05 & S/N & $\Delta_{\rm tot}$ & $RMS_{\rm obs}$ & Significance ($\sigma$)\\
\hline
$\Delta$([O/Fe])   & 0.02 & 0.05 & 0.01 & 0.01 & 0.10 & 0.11 & 0.12$\pm$0.03 & 0.3\\
$\Delta$([Na/Fe])  & 0.02 & 0.01 & 0.00 & 0.00 & 0.05 & 0.05 & 0.21$\pm$0.05 & 3.2\\
$\Delta$([Mg/Fe])  & 0.01 & 0.01 & 0.01 & 0.00 & 0.07 & 0.07 & 0.09$\pm$0.02 & 1.0\\
$\Delta$([Al/Fe])  & 0.01 & 0.04 & 0.01 & 0.01 & 0.09 & 0.10 & 0.18$\pm$0.05 & 1.6\\
$\Delta$([Ca/Fe])  & 0.05 & 0.01 & 0.00 & 0.01 & 0.07 & 0.09 & 0.09$\pm$0.02 & 0.0\\
$\Delta$([TiI/Fe]) & 0.07 & 0.02 & 0.00 & 0.01 & 0.05 & 0.09 & 0.10$\pm$0.02 & 0.5\\
$\Delta$([Cr/Fe])  & 0.06 & 0.01 & 0.01 & 0.01 & 0.04 & 0.07 & 0.06$\pm$0.01 & 1.0\\
$\Delta$([Fe/H])   & 0.03 & 0.02 & 0.01 & 0.00 & 0.02 & 0.04 & 0.08$\pm$0.03 & 1.2\\
$\Delta$([Ni/Fe])  & 0.00 & 0.02 & 0.01 & 0.01 & 0.04 & 0.05 & 0.05$\pm$0.01 & 0.0\\
$\Delta$([Ba/Fe])  & 0.04 & 0.02 & 0.03 & 0.00 & 0.08 & 0.10 & 0.11$\pm$0.03 & 0.3\\
$\Delta$([Eu/Fe])  & 0.03 & 0.06 & 0.01 & 0.01 & 0.12 & 0.14 & 0.11$\pm$0.03 & 1.0\\
\hline                                   
\end{tabular}
\label{t3}
\end{table*}

\begin{figure}
\plotone{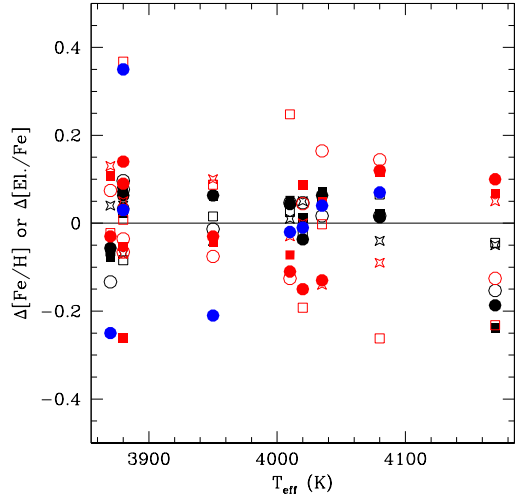}
\caption{Difference between the abundance of each star and the mean abundance of the cluster.
Different elements are reported with different symbols. BLACK: [Fe/H] (Filled circles), [Ca/Fe] (empty circles),
[Ti/Fe] (filled squares), [Cr/Fe] (empty squares), [Ni/Fe] (empty stars). RED: [Ba/Fe] (filled circles), [Eu/Fe] (empty circles),
[O/Fe] (filled squares), [Na/Fe] (empty squares), [Mg/Fe] (empry stars). BLUE: [Al/Fe] (filled circles).}
\label{fig03}
\end{figure}

As a cross check of our abundance analysis, we plot in Fig.~\ref{fig03}
for each element the difference between the abundance of each star and the mean abundance of the cluster.
We see that there is no evident trend of the abundances with temperature.
This indicates that the methodology used to obtain chemical abundances is consistent over the
entire temperature range.
A detailed internal error analysis was performed using the method described in \citet{Ma08}.
It gives us $\sigma$(T$_{eff}$)=40 K, $\sigma$(log(g))=0.15, and $\sigma$(v$_{t}$)=0.05
km/s. The error on [Fe/H] due to the S/N is 0.03 dex.
Then we choose star \#097 as representative of the sample,
varied its T$_{\rm eff}$, log(g), [Fe/H], and v$_{\rm t}$ according to
the atmospheric errors we just obtained, and redetermined the abundances.
Results are shown in Tab.~\ref{t3}, including the error due to the noise
of the spectra. This error was estimated for elements whose abundance was
obtained by EQWs, as the errors on the mean given by
MOOG, and for elements whose abundance was obtained by spectrum-synthesis, as
the error given by the fitting procedure. $\Delta_{\rm tot}$ is the
squared sum of the individual errors. For each element we report the
observed spread of the sample ($RMS_{\rm obs}$) with its error and in the final column the
significance (in units of $ \sigma$) calculated as the absolute value of the difference between 
$RMS_{\rm obs}$ and $\Delta_{\rm tot}$ divided by the error on $RMS_{\rm obs}$. 
This tells us if the observed dispersion $RMS_{\rm obs}$ is
intrinsic or due to observational errors. Values larger than 3$\sigma$ imply
an intrinsic dispersion in the species chemical abundance among
the cluster stars.

\section{Results}

\subsection{Iron-peak, $\alpha$ elements and heavy elements}

The mean iron content we obtained is:

$$[Fe/H]=-1.58\pm0.03$$

with a dispersion of:

$$\sigma_{[Fe/H]}=0.08\pm0.02$$

Reported errors are errors on the mean.
The measured iron dispersion in Tab.~\ref{t3} well agrees with the 
dispersion due to measurement errors so we no evidence for an intrinsic Fe abundance
spread. As far as other iron-peak elements are concerned, both Cr and Ni are syb-solar.

The $\alpha$ elements Ca and Ti are overabundant compared to the
Sun. This is a feature common to almost all Galactic GC and Halo field stars as well as to very metal-poor stars ([Fe/H]$<$-1.5) in outer galaxies. 
Based on these elements we derive for the cluster a mean $\alpha$ element
abundance of:

\begin{center}
$[\alpha/Fe]=+0.32\pm0.01$
\end{center}

As far as heavy elements are concerned Ba is slightly sub-solar, while Eu is super-solar.

\begin{figure}
\plotone{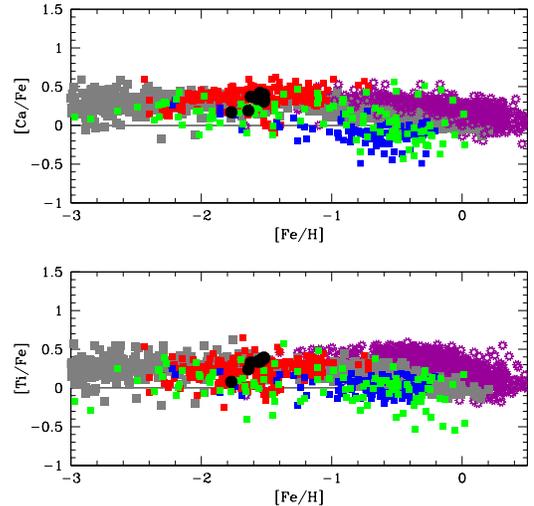}
\caption{Ca and Ti abundances for FSR 1758 (black filled circles) compared with a
   variety of Galactic and extragalactic environments: Galactic Globular Clusters (red filled squares), Disk and Halo stars (gray filled squares), Bulge stars (purple stars), Magellanic clouds (blue filled squares), and  other dwarf and ultra-faint dwarf galaxies (Draco, Sextans, Ursa Minor,  Sagittarius, Bootes I and Hercules, green filled squares). See text for more details.}
\label{fig04}
\end{figure}

Figure \ref{fig04} shows the star by
star (black filled circles) $\alpha$-element abundances of the cluster compared with a variety of galactic and extra-galactic objects. We have included values from GGCs 
\citep[red filled squares]{Ca09B,Ca10,Ca14a,Ca14b,Ca15a,Ca15b,Vi10,Vi11,Vi13,Mu13,Sa15}, Disk and Halo stars \citep[gray filled squares]{Fu00,Re03,Re06,Ca04,Si04,Ba05,Fr07,Jo12,Jo14},
the Bulge \citep[purple stars]{Go11}, and extra-galactic objects such as Magellanic
clouds \citep[blue filled squares]{Po08,Jo06,Mu08,Mu09},  
Draco, Sextans, Ursa Minor and Sagittarius dwarf galaxy and
the ultra-faint dwarf spheroidals Bo\"{o}tes I and Hercules
\citep[green filled squares]{Mo05,Sb07,Sh01,Is14,Ko08}.

The $\alpha$ elements in FSR 1758 follows the same trend as Galactic GCs and
are fully compatible with Halo and Thick Disk field stars while it falls in a region 
scarcely populated by extragalactic objects.
So, according to its  $\alpha$-element content, FSR 1758 is likely a genuine Galactic cluster.

\begin{figure}
\plotone{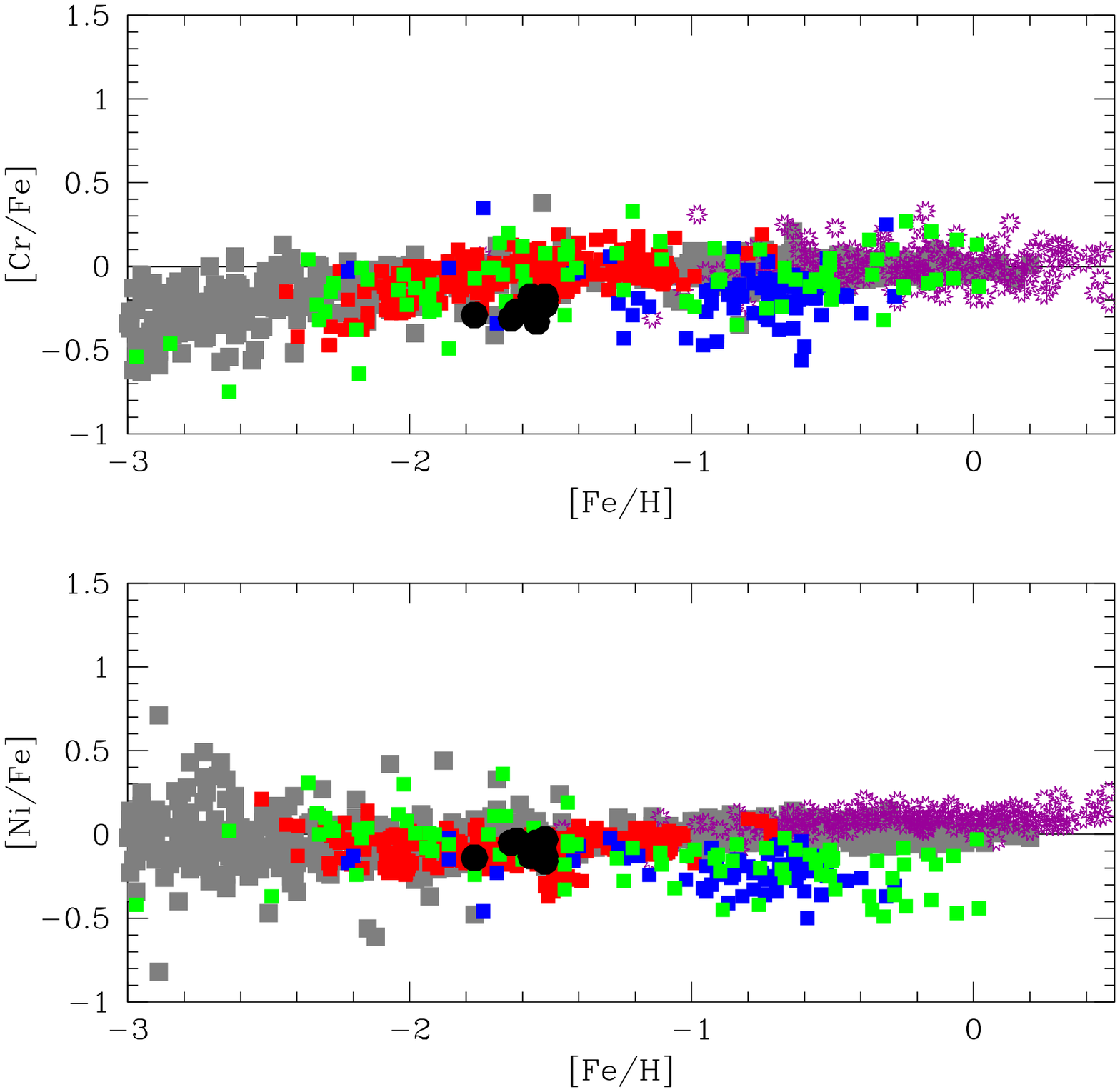}
\caption{Cr and Ni abundances for FSR 1758 (black filled circles) compared with a
   variety of Galactic and extragalactic environments: Galactic Globular Clusters (red filled squares), Disk and  Halo stars (gray filled squares), Bulge stars (purple stars), Magellanic clouds (blue filled squares), andother dwarf and ultra-faint dwarf galaxies (Draco, Sextans, Ursa Minor,  Sagittarius, Bootes I and Hercules, green filled squares). See text for more details.}
\label{fig05}
\end{figure}

\begin{figure}
\plotone{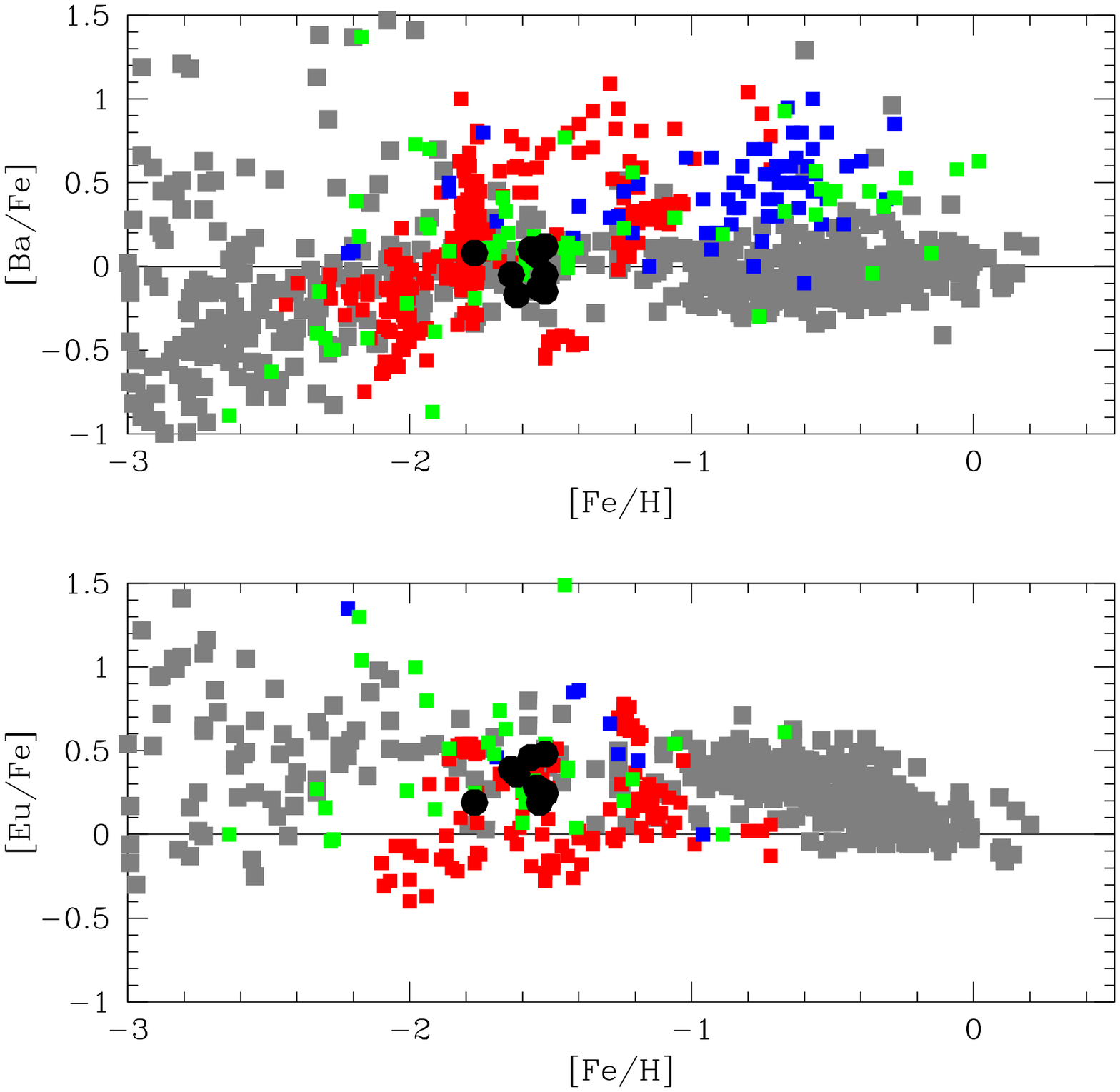}
\caption{Ba and Eu abundances for FSR 1758 (black filled circles) compared with a
   variety of Galactic and extragalactic environments: Galactic Globular Clusters (red filled squares), Disk and Halo stars (gray filled squares), Bulge stars (purple stars), Magellanic clouds (blue filled squares), and other dwarf and ultra-faint dwarf galaxies (Draco, Sextans, Ursa Minor, Sagittarius, Bootes I and Hercules, green filled squares). See text for more details.}
\label{fig06}
\end{figure}

The chemical abundances for the iron-peak elements Cr and Ni and for the heavy-elements Ba and Eu are reported in
Fig.~\ref{fig05} and ~\ref{fig06} respectively. Around FSR 1758 metallicity, Galactic and extragalactic environments share the
same abundances. We underline the low Cr content of the cluster compared with the general trend, at the very lower
boundary.

Finally for all $\alpha$, iron-peak and heavy elements Tab.~\ref{t3} shows that the
observed dispersion agrees well with the measurement errors so we can rule out
any intrinsic abundance spread. We check also for possible correlations of
these elements with lights elements such Na and Al, but we did not find
evidence for significant trends.

\begin{figure}
\centering
\includegraphics[width=8cm]{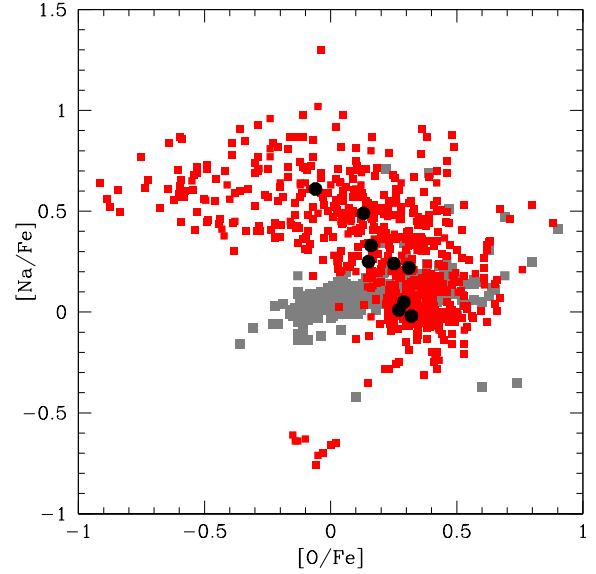}
\caption{Na-O anticorrelations for FSR 1758 (black filled circles) compared with
   Galactic Globular Clusters (red filled squares) and Disk plus Halo stars 
   (gray filled squares);  See text for more details.}
\label{fig07}
\end{figure}

\begin{figure}
\centering
\includegraphics[width=8cm]{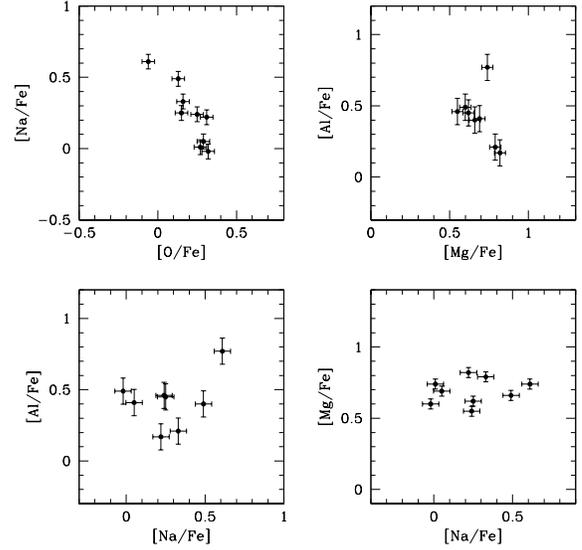}
\caption{Ligh elements correlations for FSR 1758 (black filled circles).  See text for more
   details.}
\label{fig08}
\end{figure}

\subsection{Light elements}

The light element Na has an observed spread that exceeds
the observational uncertainties (see Tab.~\ref{t3}). O, Mg and Al instead show
homogeneity, at least as far as observed dispersions are concerned.\\
In Fig.~\ref{fig07} we compare the O and Na abundances of the targets with different
environments. A clear Na-O anticorrelation appears.
The content of the two elements for the FSR 1758 first generation
stars (the targets with [O/Fe]$\sim$0.3, [Na/Fe]$\sim$0.0)
well matches the mean O and Na abundances of
the Milky Way Halo and the mean O and Na content of the other GC first
generation stars. On the other hand the cluster shows depletion in O
and Na enhancement as far as the subsequent generation targets
are concerned (the targets with [O/Fe]$<$0.3, [Na/Fe]$>$0.1).
Fig.~\ref{fig08} shows the relations between all the light elements available from our study. 
Al shows the second largest spread among the elements (0.18 dex), however it is not significant according to our error analysis. There is also no apparent relation between Na and Mg or Al at odd with wha found in other GCs (i.e. M28, \citealt{Vi17}), or a relation between Mg and Al.
Further studies with larger datasets are required to confirm or disproof this result.

\section{The orbit}

\begin{figure}
\centering
\includegraphics[width=8cm]{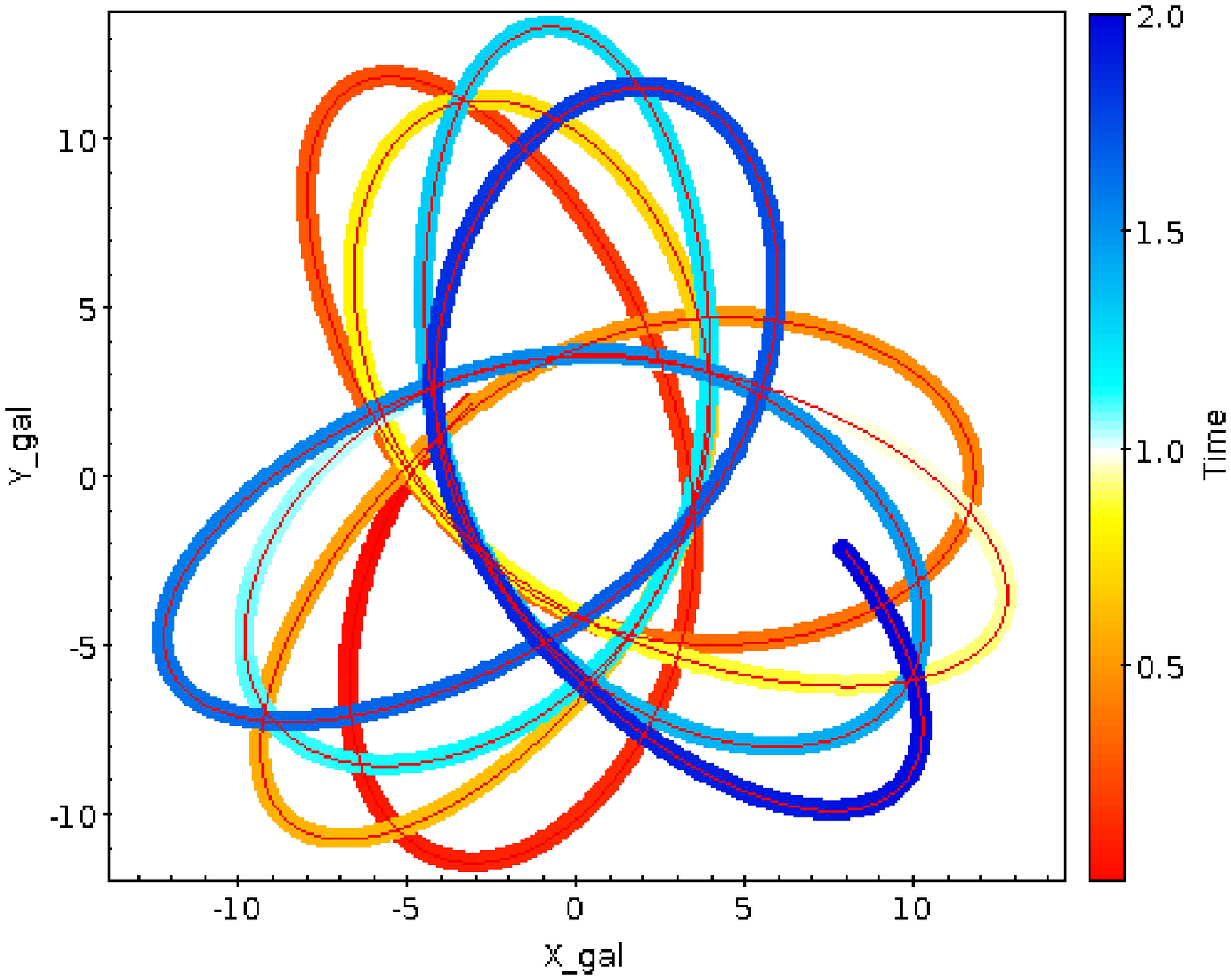}
\caption{Projecton on the X,Y galactic plane of the orbit. See text for more details.}
\label{fig09}
\end{figure}

\begin{figure}
\centering
\includegraphics[width=8cm]{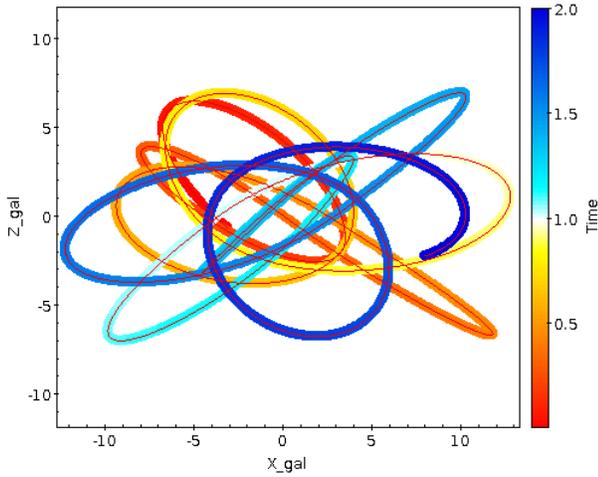}
\caption{Projecton on the X,Z galactic plane of the orbit. See text for more details.}
\label{fig10}
\end{figure}

FSR 1758 is located at 11.5 kpc from the Sun, on the other side of the Galactic Bulge.
In order to check if it belongs to the Bulge we calculated its orbit using GravPot16 \footnote{https://gravpot.utinam.cnrs.fr/} \citep{Fe17}. This is a web-code that allow to calculate orbits based on a Galactic gravitational potential driven by the Besançon Galaxy
Model mass distribution. As input parameters for the cluster we assumed:\\

RA$_{cluster}$=262.810$^0$\\

DEC$_{cluster}$=-39.815$^0$\\

Distance=11.5\ kpc\\

pm$_{RA}$=-2.7911\ mas/year\\

pm$_{DEC}$=+2.6042\ mas/year\\

RV$_{Helio}$=226.7\ km/s\\

As far as the model is concerned we used:

R$_{\odot}$= 8.2 kpc\\

U$_{\odot},V{\odot},W{\odot}$=10.0,225.2,7.2 km/s\\

We integrated to orbit forward for 2 Gyrs.
In Fig.~\ref{fig09} and Fig.~\ref{fig10} we report the projection of the orbit on the X,Y  and X,Z Galactic planes. Fig.~\ref{fig09} shows that FSR 1758 orbit does not enter the Bulge region and the cluster minimum distance from the center is about 3.8 kpc while the maximum distance is 13.6 kpc. Fig.~\ref{fig10} on the other hand suggests an orbit with a maximum height of about 7 kpc. The eccentricity and the Z component of the orbital angular momentum (L$_Z$) are 0.56 and 1.20 kpc$\times$km/s respectively.\\

We compare our orbit with that published by \citet{Si19} where the author use initial parameters very similar to ours:
\ \\

RA$_{cluster}$=262.806$^0$\\

DEC$_{cluster}$=-39.822$^0$\\

Distance=11.5\ kpc\\

pm$_{RA}$=-2.85\ mas/year\\

pm$_{DEC}$=+2.55\ mas/year\\

RV$_{Helio}$=227\ km/s\\

He find a perincenter at 3.8$^{+0.9}_{-0.9}$ kpc and an apocenter at 16$^{+8}_{-5}$ kpc.
Both parameters anicely gree with ours. Also the maximum height agrees with \citet{Si19} value beeing 7$\div$8 kpc. GravPot16 includes a Bar as far as the Bulge potential is concerned, while \citet{Si19} assumes a spherical potential for the Bulge. Since the cluster does not enter the Bulge region, the gravitational potential of the Bar does not affect its orbits significantly as the comparison of the two results shows.\\

We then support the  \citet{Si19} classification of FSR 1758 as a Halo cluster based on its orbital parameters, since both its highly eccentric orbits and the maximum distance from the Galactic plane do not match typical Disk orbits that are more circular and confined within 1 Kpc from the plane. We confirm also that, due to the L$_Z$ value we find, the orbit is retrograde.

\section{Discussion and Conclusions}

Our target stars were carefully selected to be members of the cluster based on Gaia proper motions. All the observed target stars turned out to be RV members of the cluster, giving confidence to our selection procedure. Based on the available spectra for 9 objects, we measure a velocity dispersion $\sigma_{RV}=4.9\pm 1.2$ km/s. This velocity dispersion is relatively small, typical for a globular cluster (e.g. \citealt{Pry93}).

We also confirm the mean RV measurement of \citet{Si19}, and therefore his conclusion that this cluster has a retrograde orbit. The orbital parameters we find agree nicely with theirs.

\citet{Si19} and \citet{My19} find that this GC has a strongly retrograde orbit with high eccentricity ($e \sim 0.6$). This is important because the orbit of \citet{Si19} favors the interpretation of \citet{My19} that this GC belongs to an early accretion event in the Milky Way.
\citet{My19} studies the action space and the GCs analyzed by \citet{Va19}, concluding that among the different substructures present, there was a clear signature of an individual event of accretion that was retrograde, corresponding to the Sequoia dwaf galaxy.

The orbit that we compute places this GC in a singular position with common properties with $\omega$ Centauri. It then appears that both clusters are the defining members of this  Sequoia dwarf galaxy. 

In summary, our main conclusions are: 

\begin{itemize}
\item  FSR1758 is a bonafide GC, and not the nucleus of a dwarf galaxy.
\item FSR1758 shows the typical Na-O anticorrelation common to almost all the GCs in the Galaxy.
\item Its velocity dispersion is small, typical for a globular cluster.
\item The orbit is retrograde and highly eccentric.
\item Based on the orbit, this cluster does not appear to belong to the Disk or Bulge components of the Milky Way, and is more consistent with that of a Galactic Halo building block.
\item This GC probably belonged to the Sequoia dwarf galaxy that contained at least two large GCs ($\omega$Cen and FSR1758), and was accreted in the past by the Milky Way.
\item  We suggest that the nucleus of this dwarf galaxy was $\omega$Cen, and not FSR1758 that appears to have homogeneous composition typical for a normal GC.
\item  We predict that the Sequoia dwarf galaxy may have had more GCs.
\end{itemize}

\ \\
\acknowledgments
SV gratefully acknowledges the support provided by Fondecyt
reg. n. 1170518. D.G. gratefully acknowledges support from the Chilean BASAL
Centro de Excelencia en Astrofisica y Tecnologias Afines (CATA) grant
PFB-06/2007.
This work has made use of data from the European Space Agency (ESA) mission
{\it Gaia} (\url{https://www.cosmos.esa.int/gaia}), processed by the {\it Gaia}
Data Processing and Analysis Consortium (DPAC,
\url{https://www.cosmos.esa.int/web/gaia/dpac/consortium}). Funding for the DPAC
has been provided by national institutions, in particular the institutions
participating in the {\it Gaia} Multilateral Agreement.
DM is supported by the BASAL Center for Astrophysics and Associated Technologies (CATA) through grant AFB 170002, by the Programa Iniciativa Científica Milenio grant IC120009, awarded to the Millennium Institute of Astrophysics (MAS), and by Proyecto FONDECYT No. 1170121.

\vspace{5mm}
\facilities{Magellan-Clay:Mike}

\software{MOOG \citep{Sn73}}

\end{document}